\documentclass[conference,9pt]{IEEEtran}
\IEEEoverridecommandlockouts
\usepackage{cite}
\usepackage{amsmath,amssymb,amsfonts}
\usepackage{graphicx}
\usepackage{textcomp}
\usepackage{booktabs}
\usepackage{tabularx}
\usepackage{multirow}
\usepackage{hyperref}
\usepackage{url}
\usepackage{nicefrac}
\usepackage{microtype}
\usepackage{xcolor}
\usepackage{stfloats}
\usepackage{tikz}
\usepackage[skip=3pt]{caption}
\usepackage[ruled,vlined,linesnumbered]{algorithm2e}

\usepackage{orcidlink}
\usepackage{comment}
\usepackage{braket}
\def\BibTeX{{\rm B\kern-.05em{\sc i\kern-.025em b}\kern-.08em
    T\kern-.1667em\lower.7ex\hbox{E}\kern-.125emX}}
\begin{document}

\title{QPredSGG: Hybrid Quantum Predicate Learning for Long-Tailed Scene Graph Generation}

\author{\IEEEauthorblockN{Prerana Ramkumar\textsuperscript{1}, Nouhaila Innan\textsuperscript{2,3}, Muhammad Shafique\textsuperscript{2,3}\\
\IEEEauthorblockA{
\textsuperscript{1}College of Engineering, American University of Sharjah, Sharjah, UAE\\
\textsuperscript{2}eBRAIN Lab, Division of Engineering, New York University Abu Dhabi (NYUAD), Abu Dhabi, UAE\\
\textsuperscript{3}Center for Quantum and Topological Systems (CQTS), NYUAD Research Institute, NYUAD, Abu Dhabi, UAE\\
g00100339@aus.edu, \{nouhaila.innan, muhammad.shafique\}@nyu.edu\\
}}}
\maketitle
\thispagestyle{empty}
\pagestyle{empty}
\begin{abstract}
Scene Graph Generation (SGG) requires relational reasoning over objects and their interactions, but performance is severely constrained by long-tail predicate imbalance. Classical SGG models frequently rely on dataset statistics, leading to biased predictions toward frequent relations rather than fine-grained semantic predicates. Although debiasing strategies improve mean recall, predicate classification in existing frameworks still relies on high-dimensional classical decision modules with high parameter overhead. This work introduces a resource-efficient hybrid quantum predicate classifier for SGG by replacing the classical predicate head in the Causal Features Enhancement Network (CFEN) with a Quantum Predicate Head (QP-Head) trained using class-balanced weighted cross-entropy (WCE). To the best of our knowledge, this work is among the first evaluations of a hybrid quantum architecture for scene graph predicate classification on Visual Genome 150. We systematically evaluate the impact of qubit count, encoding strategy, entangling topology, and circuit depth. Under a 4-qubit configuration using Amplitude Embedding and Strongly Entangling Layers, the QP-Head projects 4096-dimensional pair features into a compact 16-dimensional quantum input representation (a 256$\times$ input dimensionality reduction) while using only 96 trainable quantum parameters. Operating within this compressed representation space under WCE training, it achieves strong long-tail recognition, reaching a peak mR@100 of 57.25\% compared to the 41.1\% mR@100 reported in the original CFEN baseline under standard training. Protocol-matched ablations indicate that the primary advantage of the quantum head lies in extreme representational compactness rather than an accuracy gain over full-dimensional weighted classical heads: it achieves global recall within 0.19 percentage points of the 2048-dimensional WCE classical baseline while operating from a 128$\times$ smaller head feature space. Scaling to 8 qubits maintains robust long-tail performance (55.38\% mR@100 with 384 quantum parameters), while depth analysis highlights key expressibility-latency trade-offs. These results demonstrate that compact hybrid quantum predicate heads can serve as parameter-efficient, hardware-executable decision layers for long-tail relational reasoning.
\end{abstract}


\section{Introduction}

\begin{figure}[htbp]
  \centering
  \includegraphics[width=\linewidth]{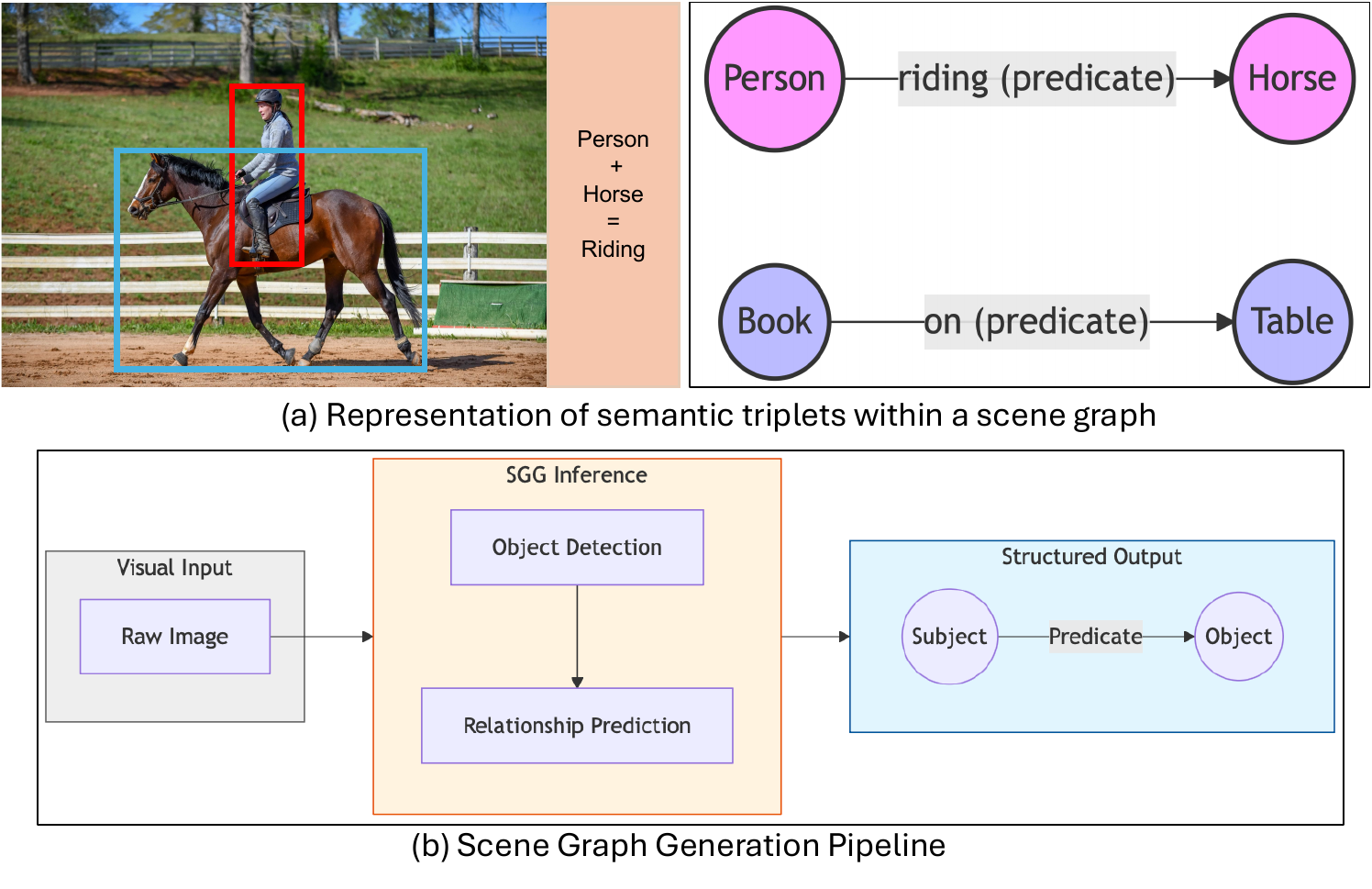}
\caption{SGG framework. (a) A semantic triplet. (b) The end-to-end pipeline from image to structured graph.}
  \label{fig:sgg-intro}
\end{figure}

Scene Graph Generation (SGG) converts an image into a structured representation in which objects are nodes and their semantic relationships are labeled directed edges~\cite{lu2016visual,hsieh2025generation,chang2021comprehensive}, expressed as triplets $\langle\textit{subject, predicate, object}\rangle$, e.g., \textit{(person, riding, horse)} (see Fig.~\ref{fig:sgg-intro}). Predicate prediction determines whether the resulting graph is semantically useful for downstream reasoning, generation, and embodied-AI tasks~\cite{anderson2018bottom,johnson2018image}.

The central obstacle to reliable predicate prediction is the long-tailed distribution of Visual Genome~\cite{krishna2017visualgenome}: a small set of generic predicates (``on'', ``of'', ``in'') dominates annotations, while semantically informative predicates are rare. Models trained with standard objectives exploit this skew and reach high global recall (R@$K$) while performing poorly under mean recall (mR@$K$), which weights all predicate classes equally. This bias dominates even strong contextual models~\cite{zellers2018neural} and is the central challenge in relational visual understanding.

Reweighting, resampling, and causal correction at inference time (Total Direct Effect, TDE~\cite{tang2020unbiased}) each reduce this bias. The Causal Features Enhancement Network (CFEN)~\cite{zhou2025cfen} separates class-specific relational evidence from class-generic scene statistics during training, improving mean recall while still mapping the resulting features through a large classical MLP head. The predicate decision module itself, independent of the training objective or the backbone, has not been substantially compressed without sacrificing long-tail recognition.

This paper addresses that gap by introducing a Quantum Predicate Head (QP-Head): a parameterized quantum circuit that replaces the classical predicate MLP and operates on a compressed relational representation produced by the CFEN backbone. The circuit encodes projected pair features into a quantum state, applies trainable entangling layers, and returns measured expectation values to a lightweight classical readout. The circuit functions as a compact decision layer inside an established SGG pipeline. We evaluate it across qubit count, encoding strategy, entangling template, and circuit depth, combining semantic metrics (R@$K$, mR@$K$) with circuit-level diagnostics (expressibility, entanglement entropy), parameter and runtime cost, and a physical-QPU feasibility test.

The quantum circuit is trained with a class-balanced weighted cross-entropy (WCE) loss on top of a linear bottleneck (4096D to 16D or 256D). To separate the contribution of the circuit from the contribution of the loss and the bottleneck, we also train the classical CFEN head under the identical WCE loss and backbone and compare it against the reference QP-Head configuration. This comparison, reported in Sec.~\ref{sec4match}, shows that the classical head matches quantum global recall and exceeds quantum mean recall while using $128\times$ more feature dimensions. The quantum head's advantage is therefore representational compactness rather than raw accuracy. All task-performance results are reported under the Predicate Classification (PredCls) setting, using ground-truth object labels and boxes.

The contributions are:
\begin{enumerate}
    \item \textbf{QP-Head architecture and design-space evaluation.} A parameterized quantum circuit that replaces the classical predicate MLP in a CFEN-based pipeline, evaluated across qubit count, encoding strategy, entangling template, and circuit depth.
    \item \textbf{Bias-aware training with a matched classical baseline.} A weighted cross-entropy protocol applied identically to the quantum head and to the classical CFEN head, enabling a loss-and-backbone-matched comparison (Sec.~\ref{sec4b}, Sec.~\ref{sec4match}).
    \item \textbf{Circuit diagnostics and computational-efficiency analysis.} Expressibility and von Neumann entropy diagnostics connected to task performance, with trainable-parameter and runtime accounting.
    \item \textbf{Physical QPU execution.} Execution of the 4-qubit QP-Head on IBM \texttt{ibm\_fez}, demonstrating stable, non-collapsed predicate predictions under real hardware noise.
\end{enumerate}

The rest of the paper is organized as follows. Section~\ref{sec2} reviews related work. Section~\ref{sec3} presents the QP-Head methodology and training protocol. Section~\ref{sec4} reports results. Section~\ref{sec5} concludes.

\section{Background and Related Work}\label{sec2}

\subsection{Scene Graph Generation and Long-Tail Debiasing}

Lu et al.~\cite{lu2016visual} formalized visual relationship detection using language priors over object co-occurrence. Zellers et al.~\cite{zellers2018neural} introduced Neural Motifs, which encode global graph context with stacked LSTMs and show that SGG models trained on Visual Genome rely heavily on dataset-level statistics rather than local visual evidence. Motifs are the standard reference point for frequency-driven predicate bias. Tang et al.~\cite{tang2019learning} proposed VCTree, which propagates relational context bidirectionally over a dynamic object tree. The CFEN backbone used follows a related BiTreeLSTM design~\cite{zhou2025cfen} and adds a causal decomposition that separates class-specific evidence from class-generic scene statistics.

Inverse-frequency reweighting and resampling reduce majority-class dominance but degrade under extreme sparsity. Tang et al.~\cite{tang2020unbiased} apply a counterfactual (TDE) correction at inference time, improving mean recall substantially at some cost to global recall. CFEN~\cite{zhou2025cfen} addresses the same bias during training through a distribution-matching loss between class-specific and class-generic features, and reports strong mean recall on PredCls; it is the classical reference backbone used throughout this work. In each of these methods, the predicate decision head that consumes the debiased representation remains a large classical MLP. This is the component this paper targets.

\subsection{Quantum Machine Learning and Circuit Diagnostics}

Quantum Machine Learning (QML) studies whether parameterized quantum circuits (PQCs) support learning through richer feature representations or more compact parameterizations~\cite{biamonte2017quantum,chang2025primer}. In the NISQ era, QML is dominated by hybrid quantum-classical models in which a PQC is a differentiable module inside a classical pipeline~\cite{preskill2018quantum,innan2024financial,innan2025lep,choudhary2025hqnn,siddiqui2025quantum,dave2025sentiqnf}, trained via Variational Quantum Algorithms~\cite{cerezo2021variational,innan2025next} using backpropagation through the simulated state vector or the parameter-shift rule on hardware.

Data encoding is a first-order design choice. Havl\'{i}\v{c}ek et al.~\cite{havlicek2019supervised} and Schuld and Killoran~\cite{schuld2019quantum} show that the encoding map determines the effective feature Hilbert space in which the model operates, which matters when high-dimensional classical features are compressed before quantum processing~\cite{innan2026spate,kashif2026design}. Circuit structure matters independently of encoding. Sim et al.~\cite{sim2019expressibility} introduced expressibility (KL divergence of the output-fidelity distribution from the Haar-random reference) and entanglement capability (von Neumann entropy of a reduced subsystem) as diagnostics for PQC design. Neither metric increases monotonically with depth~\cite{vyskubov2026scaling}, and excessive depth risks barren plateaus, where gradients vanish exponentially~\cite{mcclean2018barren}. We use both diagnostics as descriptive circuit-level indicators alongside task performance.

\subsection{Quantum Methods in Computer Vision}

Hybrid QML in vision has addressed image classification through quantum kernels, transfer learning, and hybrid networks on reduced-dimensionality inputs or small benchmarks~\cite{kharsa2023advances,senokosov2024quantum,dutta2025quiet,innan2025qnn}, reflecting current limits on qubit count and circuit depth. SGG is a substantially harder setting: it requires pairwise relational reasoning, fine-grained predicate discrimination, and operation under a severe long-tail label distribution. This work evaluates a hybrid quantum architecture for predicate classification on a large-scale relational vision benchmark, connecting SGG's context- and causality-driven debiasing methods with QML's circuit-design methodology.

\section{Methodology}\label{sec3}
\begin{figure*}[htbp]
  \centering
  \includegraphics[width=\linewidth]{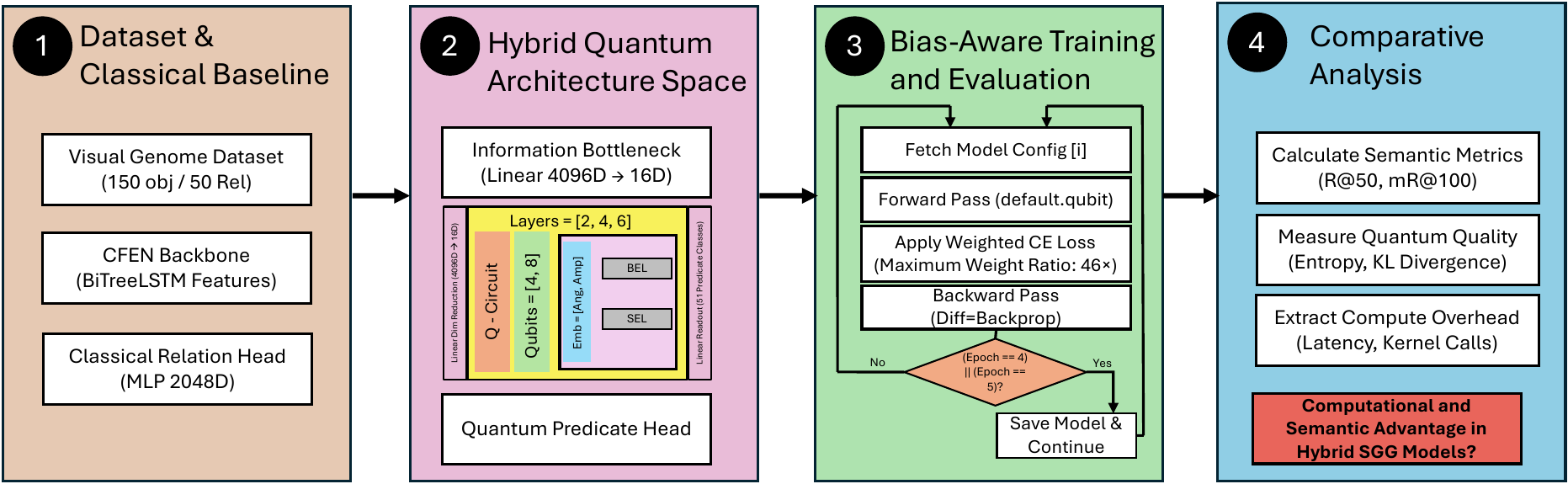}
\caption{The four-stage experimental pipeline: dataset and baseline, quantum architecture search space, bias-aware training and checkpointing, and comparative analysis.}
  \label{fig:pipeline}
\end{figure*}

The methodology follows a four-stage pipeline (Fig.~\ref{fig:pipeline}). Stage 1 establishes the Visual Genome dataset and the CFEN classical baseline. Stage 2 replaces the classical predicate head with a QP-Head and defines a controlled architecture search space. Stage 3 trains all candidates under a class-balanced objective. Stage 4 compares configurations on semantic, quantum-quality, and computational-overhead axes.

\subsection{Dataset and Classical Baseline}

Visual Genome~\cite{krishna2017visualgenome} provides densely annotated scene graphs over 150 foreground object categories and 50 foreground predicate classes, with an additional background class used internally during classification. The dataset exhibits a severely long-tailed predicate distribution: frequent spatial predicates (``on'', ``of'') dominate, while specific predicates (``flying in'', ``painted on'') are rare.

The classical backbone, CFEN, uses a Bidirectional Tree LSTM (BiTreeLSTM) to propagate relational context along the scene parse tree, producing a fused pair embedding $\mathbf{h}_{ij}\in\mathbb{R}^{4096}$ for each candidate object pair $(i,j)$. The classical predicate classifier is a 2048-dimensional MLP mapping $\mathbf{h}_{ij}$ to a distribution over $C=51$ classes, trained with standard cross-entropy:
\begin{equation}
  \mathcal{L}_{\mathrm{CE}} = -\sum_{c=1}^{C} y_c \log \hat{p}_c,
  \label{eq:ce}
\end{equation}
This configuration (C1 in Table~\ref{tab:experiments}) is the classical reference point, evaluated with Global Recall (R@$K$) and Mean Recall (mR@$K$) at $K\in\{50,100\}$.

\subsection{Hybrid Quantum Architecture Space}

The QP-Head replaces the classical MLP while reusing the same CFEN backbone, so all comparisons occur at the predicate-head level with an identical visual-context representation. A classical projection layer first maps $\mathbf{h}_{ij}\in\mathbb{R}^{4096}$ into a quantum-compatible feature vector, which is embedded into an $n$-qubit state using angle embedding (features mapped to single-qubit rotation angles) or amplitude embedding (normalized features encoded into state amplitudes). A trainable variational circuit then processes the encoded state.

\begin{figure}
    \centering
    \includegraphics[width=1\linewidth]{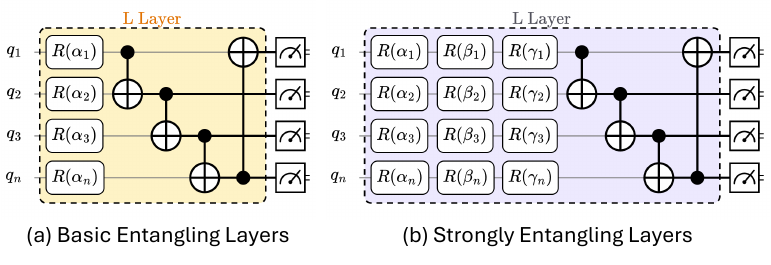}
 \caption{Representative QP-Head circuits: (a) Basic Entangling Layers, (b) Strongly Entangling Layers. Both terminate in measurements feeding the classical readout.}
    \label{fig:quantum_circuits}
\end{figure}

The search space spans qubit count $n\in\{4,8\}$, embedding $\in\{$Angle, Amplitude$\}$, circuit depth $L\in\{2,4,6\}$, and entangling template $\in\{$Basic Entangling Layers (BEL), Strongly Entangling Layers (SEL)$\}$ (Fig.~\ref{fig:quantum_circuits}). Measured expectation values form a compact quantum feature vector, passed to a lightweight classical readout that produces logits over $C=51$ classes. This design isolates how encoding, qubit count, depth, and entangling structure affect long-tail predicate classification, with the backbone and readout held fixed.

\subsection{Bias-Aware Training and Evaluation}

Every hybrid configuration is trained under the same class-balanced protocol and validation procedure, indexed by qubit count, embedding, depth, and entangling template. The forward pass uses PennyLane's \texttt{default.qubit} simulator with \texttt{diff\_method="backprop"} for differentiation through the simulated circuit.

\subsubsection{Weighted Cross-Entropy Loss}
Standard CE training under Visual Genome's long tail improves global recall while leaving mean recall low, because gradient contributions are dominated by frequent classes. We adopt Weighted Cross-Entropy (WCE):
\begin{equation}
  \mathcal{L}_{\mathrm{WCE}} =
  -\sum_{c=1}^{C} w_c\, y_c \log \hat{p}_c,
  \qquad
  w_c = \frac{1}{f_c},
  \label{eq:wce}
\end{equation}
where $f_c$ is the empirical training frequency of class $c$. Weights are clipped and normalized to a maximum rare-to-frequent ratio of $46\times$ to keep gradients stable. All QP-Head configurations reported from Sec.~\ref{sec4b} onward use WCE. We also train the classical CFEN-MLP head under the identical WCE loss and backbone, retaining its original 2048-D representation, for the matched comparison in Sec.~\ref{sec4match}.

Training (Algorithm~\ref{alg:training}) computes and clips inverse-frequency weights; for each epoch and mini-batch, extracts CFEN pair embeddings $\mathbf{h}_{ij}$, projects and encodes them into the QP-Head circuit, measures expectation values, computes $\mathcal{L}_{\mathrm{WCE}}$, backpropagates, and updates parameters with gradient clipping; every $T_{\mathrm{val}}$ epochs, it evaluates R@$K$/mR@$K$ and checkpoints without applying a single-metric selection threshold.

\begin{algorithm}[htbp]
\small
\caption{Bias-Aware QP-Head Training}
\label{alg:training}
\KwIn{Training data $\mathcal{D}$, class frequencies $\{f_c\}_{c=1}^{C}$, epochs $E$, validation interval $T_{\mathrm{val}}$}
\KwOut{Trained QP-Head parameters and validation checkpoints}
Compute $w_c=1/f_c$; clip/normalize to a $46\times$ max ratio\;
\For{$e=1$ \KwTo $E$}{
    \ForEach{mini-batch $\mathcal{B}\subset\mathcal{D}$}{
        Obtain CFEN pair embeddings $\mathbf{h}_{ij}\in\mathbb{R}^{4096}$\;
        Project and encode into the QP-Head circuit; measure expectation values and compute logits\;
        Compute $\mathcal{L}_{\mathrm{WCE}}$ (Eq.~\ref{eq:wce}); backpropagate; clip gradients and update parameters\;
    }
    \If{$e \bmod T_{\mathrm{val}} = 0$}{
        Evaluate R@$K$, mR@$K$ on validation; save checkpoint (no single-metric threshold)\;
    }
}
\end{algorithm}

This checkpointing strategy avoids selecting on a single metric. R@$K$-only selection favors frequent predicates, and mR@$K$-only selection can capture transient fluctuations. Checkpoints support a joint analysis of R@$K$, mR@$K$, and circuit-level diagnostics across training.
\subsection{Comparative Analysis}

Stage~4 of Fig.~\ref{fig:pipeline} consolidates the trained
configurations across three evaluation axes: semantic performance,
quantum quality, and computational overhead. This stage assesses whether
the QP-Head provides a trade-off relative to the classical
predicate head, particularly in terms of long-tail predicate recognition,
circuit expressivity, and execution cost.

\subsubsection{Semantic Metrics}

All models are evaluated using Global Recall and Mean Recall at
$K \in \{50,100\}$. Global Recall (R@$K$) measures the fraction of
ground-truth relationships recovered among the top-$K$ predicted
relationships, regardless of predicate class. It therefore reflects the
model's ability to recover correct relationships at the scene level but
is strongly influenced by frequent predicates. Mean Recall (mR@$K$)
computes recall independently for each predicate class and then averages
the class-wise values uniformly across all $C=51$ predicate classes.
This gives rare and frequent predicates equal weight, making mR@$K$ the
primary indicator of long-tail predicate recognition and the main metric
targeted by the weighted training objective.

\subsubsection{Quantum Quality Metrics}

To analyze the behavior of the quantum component beyond task-level
performance, two circuit-level metrics are considered: expressibility and
entanglement. \textit{Expressibility} measures how closely the output
fidelity distribution of a parameterized quantum circuit approaches the
Haar-random fidelity distribution. Following~\cite{sim2019expressibility}, it is
computed as the KL divergence between the PQC output fidelity
distribution and the Haar-random fidelity distribution:
\begin{equation}
  \mathcal{E} = D_{\mathrm{KL}}\!\left(
    \hat{F}_{\mathrm{PQC}} \,\|\, F_{\mathrm{Haar}} \right),
  \label{eq:kl}
\end{equation}
where $\hat{F}_{\mathrm{PQC}}$ denotes the empirical fidelity
distribution obtained from the sampled circuit outputs and
$F_{\mathrm{Haar}}$ denotes the reference Haar fidelity distribution.
Lower values of $\mathcal{E}$ indicate that the sampled circuit states
more closely match the Haar-random fidelity distribution. However, high expressibility alone does not guarantee better classification performance.

\textit{Entanglement} is measured using the von Neumann entropy of a
subsystem $A$, obtained by tracing out the remaining qubits:
\begin{equation}
  S_A = -\mathrm{Tr}(\rho_A \log \rho_A),
  \qquad
  \rho_A = \mathrm{Tr}_{\bar{A}}(\rho),
  \label{eq:entanglement_entropy}
\end{equation}
where $\log$ denotes the natural logarithm. Higher $S_A$ indicates stronger bipartite entanglement between subsystem
$A$ and the rest of the quantum register. This metric is used to examine
whether deeper or more strongly connected circuits generate richer
quantum correlations, and whether these correlations align with improved
semantic performance.

\subsubsection{Computational Overhead}

Computational overhead is analyzed through parameter count and inference
latency. Parameter counts are decomposed into classical and quantum
components to assess whether the QP-Head reduces the number of trainable
parameters relative to the classical predicate MLP. Inference latency is
profiled using the PyTorch autograd profiler, with total CUDA execution
time reported alongside kernel-level fragmentation statistics, including
\texttt{aten::mul}, \texttt{aten::einsum}, and \texttt{aten::bmm}. These
measurements are examined as a function of qubit count, circuit depth,
and entangling template.

By combining semantic metrics, circuit-level diagnostics, and execution
cost measurements, the comparative analysis evaluates whether the
QP-Head offers a favorable balance between long-tail recognition,
quantum representational properties, and computational cost for
scene graph generation.

\section{Results and Discussion}\label{sec4}

\subsection{Experimental Setup}

Experiments proceed in three phases (Table~\ref{tab:experiments}). Phase I reproduces the CFEN reference and measures the effect of WCE against standard CE. Phase II searches the 4-qubit encoding and entangling space. Phase III scales to 8 qubits and ablates circuit depth. All experiments run on Kaggle GPU environments; hyperparameters are listed in Table~\ref{tab:hyperparam}. Classical and quantum-hybrid configurations exhibit different training dynamics (they converge at different rates). We therefore report the best recorded value for each stated metric, with the corresponding epoch provided where available, rather than restricting the comparison to epoch-matched snapshots.
\begin{table}[htbp]
\caption{Experimental Setup and Hyperparameters}
\label{tab:hyperparam}
\begin{center}
\begin{tabular}{llc}
\toprule
\textbf{Category} & \textbf{Parameter} & \textbf{Value} \\
\midrule
Dataset & Target Dataset & VG150 \\
& Object / Relation Classes & 151 / 51 \\
\addlinespace
Model Architecture & Feature Dimension ($d_{feat}$) & 1024 \\
& Feature Fusion Strategy & Summation \\
& DM Loss Weight ($\lambda_{DM}$) & 0.4 \\
\addlinespace
Optimization & Batch Size & 128 \\
& Training Epochs & 56 \\
& Base Learning Rate ($\eta$) & 0.001 \\
& Weight Decay & 0.0001 \\
& Optimizer Momentum & 0.9 \\
\addlinespace
Reproducibility & Random Seed & 1337 \\
\bottomrule
\end{tabular}
\end{center}
\end{table}

\begin{table*}[h]
\centering
\caption{Experimental configurations for the classical CFEN baseline and hybrid QP-Head models.}
\label{tab:experiments}
\resizebox{\textwidth}{!}{%
\begin{tabular}{@{}llccccccc@{}}
\toprule
\textbf{Exp ID} & \textbf{Model Architecture} & \textbf{Loss Function} & \textbf{Qubits} & \textbf{Encoding Strategy} & \textbf{Entangling Strategy} & \textbf{Layers} & \textbf{Classical Params} & \textbf{Quantum Params} \\
\midrule
\multicolumn{9}{c}{\textit{Phase I: Classical Baselines and Loss Ablation}} \\
\midrule
\textbf{C1} & Classical CFEN (MLP) & Standard CE & -- & -- & -- & -- & 77,748,276 & 0 \\
\textbf{C2} & Classical CFEN (MLP) & Weighted CE & -- & -- & -- & -- & 77,748,276 & 0 \\
\midrule
\multicolumn{9}{c}{\textit{Phase II: 4-Qubit QP-Head Search}} \\
\midrule
\textbf{Q4a} & Hybrid QP-Head & Weighted CE & 4 & Angle & Basic (BEL) & 2 & 69,516,196 & 96 \\
\textbf{Q4b} & Hybrid QP-Head & Weighted CE & 4 & Angle & Strongly Entangling (SEL) & 2 & 69,516,196 & 96 \\
\textbf{Q4c} & Hybrid QP-Head & Weighted CE & 4 & Amplitude & Strongly Entangling (SEL) & 2 & 69,516,196 & 96 \\
\midrule
\multicolumn{9}{c}{\textit{Phase III: 8-Qubit Scaling and Depth Ablation}} \\
\midrule
\textbf{Q8a} & Hybrid QP-Head & Weighted CE & 8 & Amplitude & Strongly Entangling (SEL) & 2 & 73,450,132 & 192 \\
\textbf{Q8b} & Hybrid QP-Head & Weighted CE & 8 & Amplitude & Strongly Entangling (SEL) & 4 & 73,450,132 & 384 \\
\textbf{Q8c} & Hybrid QP-Head & Weighted CE & 8 & Amplitude & Strongly Entangling (SEL) & 6 & 73,450,132 & 576 \\
\bottomrule
\end{tabular}%
}
{\raggedright \small \textit{Note:} WCE applies class-dependent weights with a maximum rare-to-frequent class weight ratio of $46\times$. Quantum parameters refer only to trainable parameters inside the PQC before the final classical readout layer. C2 is compared against the reference QP-Head in Sec.~\ref{sec4match}.\par}
\end{table*}

\subsection{Effect of Class-Balanced Training on the QP-Head}\label{sec4b}
\begin{figure*}[htpb]
    \centering
    \includegraphics[width=1\linewidth]{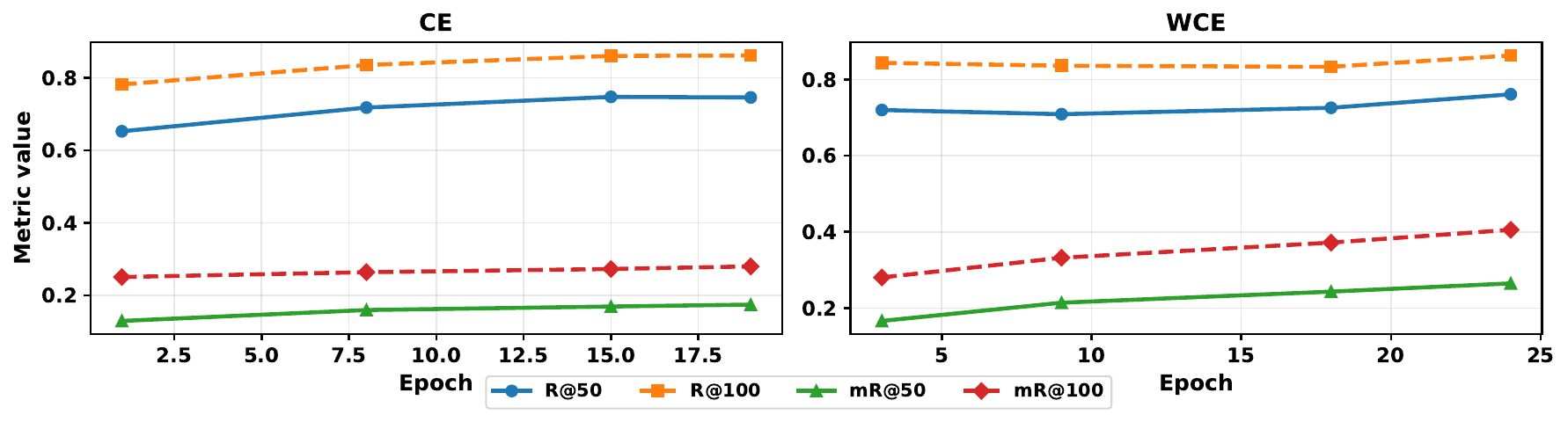}
\caption{4-qubit QP-Head training dynamics under CE (left) versus WCE (right), Angle Embedding with Basic Entangling Layers. WCE raises mR@50 and mR@100 while R@50 remains stable.}
    \label{fig:ce_wce_training_dynamics}
\end{figure*}

Fig.~\ref{fig:ce_wce_training_dynamics} compares the same 4-qubit QP-Head (Angle Embedding, Basic Entangling Layers) trained under CE and WCE. Under CE, R@50 rises from 0.6526 (Epoch~1) to 0.7460 (Epoch~19), while mR@50 plateaus at 0.1748: global recall improves while rare-predicate recognition does not, consistent with the frequency-driven bias reported for classical SGG models~\cite{zellers2018neural}.

Under WCE, mR@50 rises from 0.1660 (Epoch~3) to 0.2647 (Epoch~24), 8.99 points above the best CE value, and mR@100 rises from 0.2802 to 0.4054, while R@50 remains stable at 0.7614. In this experiment, class-balanced training improves the best recorded long-tail metrics while maintaining comparable global recall. Accordingly, all subsequent QP-Head configurations are trained using WCE.

\subsection{Protocol-Matched Comparison of the Classical and Quantum Heads under WCE}\label{sec4match}

To isolate the contribution of the quantum circuit from the contribution of the WCE objective, we train the classical CFEN predicate head (C2 in Table~\ref{tab:experiments}: the original 2048-D MLP, no bottleneck) under the identical WCE loss and CFEN backbone, and compare it against the reference 4-qubit QP-Head (Angle Embedding with Basic Entangling Layers, the configuration profiled in Fig.~\ref{fig:ce_wce_training_dynamics}) at each model's converged checkpoint. Table~\ref{tab:matched_comparison} reports the result.

\begin{table}[htbp]
\centering
\caption{WCE-matched comparison: classical CFEN-MLP head versus reference 4-qubit QP-Head (Angle Embedding, Basic Entangling), same backbone, same loss, same evaluation protocol.}
\label{tab:matched_comparison}
\begin{tabular}{lccl}
\toprule
\textbf{Metric} & \textbf{Classical} & \textbf{Quantum} & \textbf{Note} \\
\midrule
Epochs to converge & $\sim$8 & $\sim$18 & Classical converges faster \\
R@50 & 72.78\% & 72.59\% & Difference: 0.19 pts \\
mR@50 & 28.99\% & 24.30\% & Classical leads by 4.7\,pts \\
Feature dimension & 2048 & 16 & Quantum is $128\times$ smaller \\
\bottomrule
\end{tabular}
\end{table}

Under an identical backbone and WCE objective, the classical and quantum heads obtain numerically similar global recall values of 72.78\% and 72.59\%, respectively. The classical head achieves higher mean recall (28.99\% versus 24.30\%) and converges roughly twice as fast. The quantum head achieves global recall within 0.19 percentage points of the classical head and approaches its mean recall while using a 128$\times$ lower-dimensional quantum-compatible input representation. The PQC itself contains 96 trainable parameters, excluding the classical backbone, projection, and readout. At this reference configuration, the quantum circuit therefore provides an advantage in representational compactness rather than accuracy.

This result defines the scope of the paper's claim: a heavily compressed quantum decision layer achieves global recall close to the classical baseline and lower, but still competitive, mean recall while operating on a quantum-compatible input representation with a 128× lower-dimensional representation. This trade-off is relevant when decision-head footprint or physical-hardware execution cost (Sec.~\ref{sec4hw}) is an important deployment constraint.

Two limitations should be considered when interpreting this comparison. First, the classical head retains its full 2048-dimensional representation, whereas the quantum model operates on a 16-dimensional input representation. Consequently, the observed performance difference cannot be attributed solely to the circuit architecture. A classical control using the same 16-dimensional bottleneck would help separate the effects of dimensional compression and circuit design. Second, the comparison uses the Angle Embedding and Basic Entangling reference configuration rather than the best-performing Amplitude Embedding and Strongly Entangling configuration identified in Sec.~\ref{sec4c}.

\subsection{4-Qubit QP-Head Search}\label{sec4c}
Holding WCE fixed, we compare Angle with Basic Entangling, Angle with Strongly Entangling, and Amplitude Embedding with Strongly Entangling (Table~\ref{tab:4qubit_summary}, best checkpoint per metric).

\begin{table*}[htbp]
  \centering
  \caption{Best-checkpoint performance of 4-qubit QP-Head configurations under WCE loss (epoch in parentheses).}
  \label{tab:4qubit_summary}
  \begin{tabular}{lccl}
    \toprule
    \textbf{Configuration} & \textbf{Best R@50} & \textbf{Best mR@100} & \textbf{Observation} \\
    \midrule
    Angle + Basic Entangling
      & 0.7960 (Ep.~48) & 0.4980 (Ep.~52) & Reference 4-qubit setting \\
    Angle + Strongly Entangling
      & 0.8050 (Ep.~48) & 0.4929 (Ep.~52) & Higher global recall \\
    Amplitude + Strongly Entangling
      & \textbf{0.8458} (Ep.~36)
      & \textbf{0.5725} (Ep.~52)
      & Best overall \\
    \bottomrule
  \end{tabular}
\end{table*}

Switching from Basic to Strongly Entangling Layers raises R@50 from 0.7960 to 0.8050 with mR@100 essentially unchanged: the stronger entangling structure improves global recovery without a long-tail cost. The larger gain comes from switching Angle to Amplitude embedding under the same entangling template: R@50 rises to 0.8458 and mR@100 to 0.5725. Amplitude encoding uses the 16-dimensional quantum state space more effectively than angle encoding under this compression ratio. This configuration reaches $S_A=0.9817$ and $\mathcal{E}=0.0354$, indicating strong entanglement and low divergence from the Haar reference. Amplitude Embedding with Strongly Entangling Layers is the primary 4-qubit configuration used in the remaining analysis.

\subsection{Scaling to 8 Qubits and Depth Ablation}\label{sec4d}
\begin{figure}[htpb]
    \centering
    \includegraphics[width=1\linewidth]{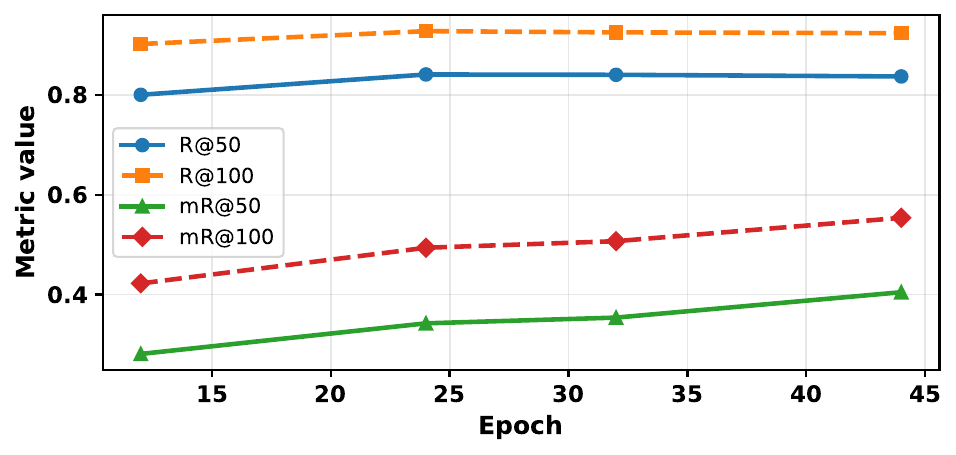}
 \caption{8-qubit QP-Head training dynamics (Amplitude Embedding, 4-layer SEL): stable global recall with improving long-tail recall.}
    \label{fig:8qubit_dynamics}
\end{figure}

Scaling the selected 4-qubit design to 8 qubits expands the encoded state space from $2^4{=}16$ to $2^8{=}256$ dimensions. Fig.~\ref{fig:8qubit_dynamics} shows mR@50 rising from 0.2807 (Epoch~12) to 0.4045 (Epoch~44) and mR@100 rising from 0.4222 to 0.5538, while R@50 stays above 0.80 and R@100 stays above 0.90 throughout. The 8-qubit configuration shows improving long-tail metrics over training while maintaining high global recall.

Table~\ref{tab:depth_ablation} ablates depth $L\in\{2,4,6\}$ at 8 qubits. KL divergence falls from 0.1116 ($L{=}2$) to 0.0432 ($L{=}4$) to 0.0293 ($L{=}6$): deeper circuits track the Haar reference more closely, but CUDA inference time rises from 214.59\,ms to 474.67\,ms and \texttt{aten::mul} calls rise from 752 to 2{,}160. Entanglement entropy decreases with depth (2.4606 to 2.3142 to 2.2690), so expressibility and entanglement do not improve together; depth buys expressibility at increasing runtime cost, not stronger entanglement. The 6-layer circuit reaches the lowest divergence, but the 4-layer circuit captures most of the expressibility gain over $L{=}2$ at substantially lower latency than $L{=}6$. We adopt the 8-qubit, 4-layer configuration as the primary scaled model.

\begin{table*}[htbp]
  \centering
  \caption{Depth ablation for the 8-qubit QP-Head (Amplitude Embedding, Strongly Entangling Layers): quantum-quality metrics and GPU inference statistics.}
  \label{tab:depth_ablation}
  \begin{tabular}{lcccl}
    \toprule
    \textbf{Metric} & \textbf{2-Layer} & \textbf{4-Layer} & \textbf{6-Layer} & \textbf{Trend} \\
    \midrule
    Quantum Parameters & 192 & 384 & 576 & Linear increase \\
    KL Div.\ from Haar ($\mathcal{E}$, $\downarrow$ better) & 0.1116 & 0.0432 & \textbf{0.0293} & Lower with depth \\
    von Neumann Entropy ($S_A$, $\uparrow$ better) & \textbf{2.4606} & 2.3142 & 2.2690 & Slight decrease with depth \\
    CUDA Inference Time & 214.59\,ms & 341.01\,ms & 474.67\,ms & Increases with depth \\
    \texttt{aten::mul} Kernel Calls & 752 & 1{,}456 & 2{,}160 & Growing fragmentation \\
    \texttt{aten::einsum} Calls & 64 & 128 & 192 & Linear scaling \\
    \bottomrule
  \end{tabular}
\end{table*}

\subsection{Parameter Efficiency and Runtime Overhead}\label{sec4e}
Table~\ref{tab:params} decomposes classical and quantum parameters for the two primary configurations. The quantum circuit contributes 96 (4-qubit) or 384 (8-qubit, 4-layer) trainable parameters, under 0.001\% of the total hybrid model in both cases, while the CFEN backbone and classical projection account for the remainder. The quantum component changes the final decision stage without materially increasing total trainable capacity: the increase in encoded state-space dimensionality (16D to 256D) is not matched by a proportional increase in trainable weights.

\begin{table*}[htbp]
  \centering
  \caption{Parameter decomposition of the primary QP-Head configurations.}
  \label{tab:params}
  \begin{tabular}{lrrrr}
    \toprule
    \textbf{Model} & \textbf{Total Params} & \textbf{Classical Params} & \textbf{Quantum Params} & \textbf{Q \%} \\
    \midrule
    4-Qubit (Amplitude + SEL, 2L)
      & 69{,}516{,}292 & 69{,}516{,}196 & 96  & 0.0001\% \\
    8-Qubit (Amplitude + SEL, 4L)
      & 73{,}450{,}516 & 73{,}450{,}132 & 384 & 0.0005\% \\
    \bottomrule
  \end{tabular}
\end{table*}

Within the 8-qubit depth ablation, runtime overhead increases with circuit depth. CUDA latency rises from 214.59\,ms to 474.67\,ms and \texttt{aten::mul} calls rise from 752 to 2{,}160 between $L{=}2$ and $L{=}6$ (Table~\ref{tab:depth_ablation}), confirming that depth, not trainable parameter count, is the main cost driver. The 8-qubit, 4-layer model is a balanced scaled configuration, and the 4-qubit, 2-layer model is the most hardware-friendly configuration.

\subsection{Hardware Feasibility Check}\label{sec4hw}
We execute the 4-qubit, 2-layer Amplitude Embedding with SEL configuration on \texttt{ibm\_fez} (IBM Heron~r2, 156 qubits, CZ native two-qubit gate, 1{,}024 shots per circuit) for 9 VG-150 validation triplets, submitted independently. This is a feasibility check on a small batch, not a hardware benchmark. Table~\ref{tab:hardware} reports the outcome: 6 of 9 correct (66.67\% batch accuracy), 1.42\,s end-to-end latency per triplet (an upper bound, since batched submission was not used), and predictions spanning four distinct predicate classes rather than collapsing to one. This is the operationally meaningful result: the trained circuit retains class-discriminative structure after transpilation and hardware noise, and the most frequent output (``Wearing'', a human-object-interaction predicate) is not simply the majority training class.

\begin{table}[htbp]
  \centering
  \caption{Physical QPU execution on \texttt{ibm\_fez} for 9 VG-150 validation triplets (4-qubit, 2-layer SEL, 96 quantum params).}
  \label{tab:hardware}
\begin{tabularx}{\columnwidth}{l l X}
    \toprule
    \textbf{Metric} & \textbf{Value} & \textbf{Notes} \\
    \midrule
    Batch Accuracy & 66.67\% & 6 of 9 correct \\
    Total Latency & 1.42\,s/triplet & Unbatched, upper bound \\
    Predicted Classes & 4 distinct & Wearing (4), Of (3), On (1), Has (1) \\
    \bottomrule
  \end{tabularx}
\end{table}

\subsection{Comparison with Existing SGG Methods}\label{sec4h}
Table~\ref{tab:sgg_comparison} places the QP-Head alongside Motifs, VCTree-TDE, and CFEN under PredCls, using each method's own published numbers. Motifs reaches high global recall but low mean recall, reflecting frequency bias. VCTree-TDE raises mean recall through causal correction at the cost of global recall. CFEN improves long-tail recall while remaining competitive on global recall. The best-tuned QP-Head variants (Amplitude Embedding with Strongly Entangling Layers, Sec.~\ref{sec4c}) reach mR@100 of 57.25\% (4-qubit) and 55.38\% (8-qubit, 4-layer), above the literature CFEN figure of 41.1\%, using 96 and 384 quantum parameters, respectively.

\begin{table*}[h]
  \centering
\caption{Comparison with representative SGG methods on VG-150 under the PredCls setting. Results for Motifs, VCTree-TDE, and CFEN are included as reported benchmarks and were obtained under experimental settings different from ours.}
  \label{tab:sgg_comparison}
  \begin{tabular}{lccccccc}
    \toprule
    \textbf{Method} & \textbf{Head Type} & \textbf{R@50} & \textbf{R@100} & \textbf{mR@50} & \textbf{mR@100} & \textbf{Classical Params} & \textbf{Quantum (PQC) Params} \\
    \midrule
    Motifs~\cite{zellers2018neural} & Classical & -- & 67.1 & -- & 15.8 & -- & 0 \\
    VCTree-TDE~\cite{tang2020unbiased} & Classical + TDE & -- & 51.6 & -- & 28.7 & -- & 0 \\
    CFEN~\cite{zhou2025cfen} & Classical & -- & -- & -- & 41.1 & 77,748,276 & 0 \\
    QP-Head 4q (Amp. + SEL) & Hybrid Quantum & 84.58 & -- & -- & 57.25 & 69,516,196 & 96 \\
    QP-Head 8q (Amp. + SEL, 4L) & Hybrid Quantum & 83.73 & 92.41 & 40.45 & 55.38 & 73,450,132 & 384 \\
    \bottomrule
  \end{tabular}
\end{table*}

The two comparisons in this section serve different purposes and should not be conflated. Table~\ref{tab:sgg_comparison} reports external context under each method's own training protocol and reports the best-tuned QP-Head configuration. Table~\ref{tab:matched_comparison} reports an internal, protocol-matched comparison using the reference QP-Head configuration and our own WCE-trained CFEN reproduction. The internal comparison in Sec.~\ref{sec4match} is the stronger evidence for evaluating the quantum circuit's contribution, because both heads share the same backbone, loss, and evaluation protocol. Whether the best-tuned quantum configuration widens the margin over our matched classical baseline is an extension of Sec.~\ref{sec4match}.

\subsection{Discussion}

The protocol-matched comparison in Sec.~\ref{sec4match} establishes the central empirical success of this work: demonstrating that a highly compact, hardware-executable quantum circuit can successfully substitute for a large classical decision head without structural performance collapse. Under an identical CFEN backbone and a matched WCE loss, the reference 4-qubit QP-Head achieves global recall close to the classical baseline while operating on a $128\times$ smaller feature representation and using only 96 trainable weights. This representational-compactness advantage is systematically driven by three key quantum design factors: highly dense amplitude-based encoding under severe dimensional compression (Sec.~\ref{sec4c}), an optimized entangling topology that yields strong relational expressibility without inducing circuit depth overhead (Sec.~\ref{sec4d}), and an ultra-lean footprint that maintains the quantum component below 0.001\% of total model parameters even when scaling up to 8 qubits (Sec.~\ref{sec4e}). Our results support the core contribution: the QP-Head substantially reduces the dimensionality of the decision-stage representation and quantum parameter count while maintaining global recall close to the matched classical baseline.

Rather than representing rigid bounds, the specific evaluation parameters of this study highlight the untapped upside and scaling potential of the hybrid quantum framework. First, the current matched baseline comparison gives the classical head an unbottlenecked advantage by allowing it to retain its full 2048-D representation space; evaluating a classical head strictly bottlenecked to the identical 16 dimensions alongside our optimized amplitude-embedded configuration will further sharpen this comparison, highlighting the effects of dimensional compression and circuit architecture to be evaluated under matched constraints.

Furthermore, our internal ablations successfully treat circuit expressibility and entanglement as actionable diagnostic tools to map out the structural trade-offs between mathematical depth and mR@$K$ metrics. While our external comparisons in Table~\ref{tab:sgg_comparison} and hardware validation in Sec.~\ref{sec4hw} serve to demonstrate broader competitive feasibility and real-world execution, they pave a clear path for future extensions into full Scene Graph Detection (SGDet) pipelines and hardware-accelerated batch deployment. Ultimately, this work delivers a proof-of-concept for quantum-enhanced visual reasoning: showing that a 96-to-384 parameter quantum decision layer can absorb high-dimensional scene contexts, reduce the dimensionality of the decision-stage representation, and reliably execute complex relational classification on physical quantum hardware.

\section{Conclusion}\label{sec5}

This paper presents the Quantum Predicate Head (QP-Head), a parameterized quantum circuit that replaces the classical predicate classifier in a CFEN-based scene graph generation pipeline, operating on compressed relational pair embeddings with a lightweight classical readout. Class-balanced (WCE) training improves long-tail recognition on the quantum head. Amplitude Embedding with Strongly Entangling Layers is the strongest compact 4-qubit configuration. Scaling to 8 qubits maintains high global recall and strong mean recall, at the cost of runtime that increases with circuit depth. A hardware run on IBM \texttt{ibm\_fez} confirms that the trained 4-qubit circuit does not collapse to a single output class under real superconducting noise.

A classical CFEN head trained under the identical WCE loss and backbone converges faster and reaches higher mean recall than the reference quantum head, while the quantum head matches global recall from a $128\times$ smaller feature representation and 96 trainable quantum parameters. The QP-Head's demonstrated advantage is representational compactness: near-classical performance from a substantially compressed representation and a small trainable circuit, executable on physical quantum hardware. Evaluating a classical head bottlenecked to the identical 16 dimensions alongside our optimized amplitude-embedded quantum configuration will further sharpen this comparison, allowing the effects of dimensional compression and circuit architecture to be evaluated under matched constraints. Extending evaluation beyond PredCls and studying noise-aware training for deeper circuits are natural next steps. The result established in this study is that a sub-400-parameter quantum circuit functions as a hardware-executable decision layer that reaches competitive mean-recall performance from a radically smaller representation, with design axes, encoding, entangling structure, and depth that behave predictably enough to be optimized under a fixed training protocol.

\section*{Code Availability}
The code developed for this paper is publicly available at \url{https://github.com/eBrain4Everyone/QPredSGG}.
\section*{Acknowledgment}
 This work was supported in part by the NYUAD Center for Quantum and Topological Systems (CQTS), funded by Tamkeen under the NYUAD Research Institute grant CG008.

\bibliographystyle{IEEEtran}
\bibliography{refs}

\end{document}